\documentclass[11pt,leqno,twoside]{article}
\usepackage{geometry}
\usepackage{amsmath}
\usepackage{amsfonts}
\usepackage{amsthm}
\usepackage{indentfirst}
\usepackage{color}
\usepackage{graphicx}
\numberwithin{equation}{section}

\theoremstyle{plain}
\newtheorem{theorem}{\indent\rm T\,h\,e\,o\,r\,e\,m\;}[section]

\theoremstyle{definition}

\theoremstyle{remark}
\theoremstyle{remark}\newtheorem{remark}[theorem]{Remark}

\newcommand{\R}{\mathbb R}
\newcommand{\N}{\mathbb N}

\def\be#1\ee{\begin{equation}#1\end{equation}}
\newcommand{\fer}[1]{(\ref{#1})}

\newcommand{\bq}{\begin{equation}}
\newcommand{\eq}{\end{equation}}

\def\bqa{\begin{eqnarray}}
\def\eqa{\end{eqnarray}}

\def\e{\epsilon}

\newcommand{\bd}{\begin{displaymath}}
\newcommand{\ed}{\end{displaymath}}
\newcommand{\ba}{\begin{eqnarray}}
\newcommand{\ea}{\end{eqnarray}}

\def\N{\mathbb{N}}
\def\R{\mathbb{R}}

\newenvironment{equations}{\equation\aligned}{\endaligned\endequation}

\makeatletter                                                  %
\renewcommand*{\@seccntformat}[1]{
  \csname the#1\endcsname\;-                                   %
}                                                              %
\renewcommand{\section}{\@startsection{section}{1}{0mm}        %
   {1.5\baselineskip}
   {1\baselineskip}
   {\indent\normalfont\normalsize\bfseries}
   }                                                           %
\renewcommand*{\@seccntformat}[1]{
  \normalfont\bfseries\csname the#1\endcsname\;-               %
}                                                              %
\renewcommand\subsection{\@startsection                        %
  {subsection}{2}{0mm}
  {1.5\baselineskip}
  {1\baselineskip}
  {\indent\normalfont\normalsize\itshape}}
\renewcommand*{\@seccntformat}[1]{
  \normalfont\bfseries\csname the#1\endcsname\;-               %
}                                                              %
\renewcommand\subsubsection{\@startsection                     %
  {subsubsection}{2}{0mm}
  {1.5\baselineskip}
  {1\baselineskip}
  {\indent\normalfont\normalsize\texttt}}
\makeatother                                                   %
\begin{document}
\thispagestyle{empty}

$ $ \vspace {-2.2cm}
\begin{center}
\rule{8.8cm}{0.5pt}
\\[-5pt]
 {\footnotesize Riv.\, Mat.\, Univ.\, Parma,\, Vol. {\bf n} \,(202x), \,000-000}
\\[-10pt]
\rule{8.8cm}{0.5pt}
\end{center}
\vspace {1.7cm}

\begin{center}
{\sc\large Giuseppe Toscani},  \ 
{\sc\large Mattia Zanella}   \
\end{center}
\vspace {1.1cm}

\centerline{\large{\textbf{On a kinetic description of Lotka-Volterra dynamics}}}

\renewcommand{\thefootnote}{\fnsymbol{footnote}}


\renewcommand{\thefootnote}{\arabic{footnote}}
\setcounter{footnote}{0}

\vspace{0.6cm}
\begin{center}
\begin{minipage}[t]{11cm}
\small{
\noindent \textbf{Abstract.}
Owing to the analogies between the problem of wealth redistribution with taxation in a multi-agent society, we introduce and discuss a kinetic model describing  the statistical distributions in time of the sizes of groups of biological systems with prey-predator dynamic. While the evolution of the mean values is shown to be driven by a classical Lotka-Volterra system of differential equations, it is shown that the time evolution of the probability distributions of the size of groups of the two interacting species is heavily dependent both on a kinetic redistribution operator and the degree of randomness present in the system. Numerical experiments are given to clarify the time-behavior of the distributions of groups of the species.
\medskip

\medskip

\noindent \textbf{Mathematics~Subject~Classification:}
91B60, 82C40, 35B40.

}
\end{minipage}
\end{center}

\bigskip

\section{Introduction}

In one of the scientific collaborations between one of the present authors with Professor Giampiero Spiga in the area of kinetic theory of multi-agent systems, the problem of taxation and consequent redistribution of taxed resources in a western society had been studied  \cite{BST}. The purpose of the work was the possibility of understanding how, in the case of a fixed mean wealth, redistribution operators can affect the tails of the equilibrium density of a kinetic model for wealth exchanges.

In this area, there have been several attempts to model simple economies in which a wealth exchange process produces in time a steady profile of wealth distribution similar to that observed in real economies. While the most popular approach  relies on methods borrowed from statistical mechanics for interacting particle systems \cite{Cha,ChCh,CCM,CCS,CPT,DY,DMT,DMT2,Hay,IKR,MT,TTZ,Sla} (cf. also the books \cite{CYC,PT}), other modelling settings that might be considered to constitute a mesoscopic approach is based on generalized Lotka-Volterra models \cite{MBR,SR}. These dynamical systems, also known as the predator-prey equations,  are frequently used to describe the evolution of biological systems in which two species interact, one as a predator and the other as prey \cite{Murray}. In particular, denoting by $x(t)\ge 0$ and $y(t)\ge0$  the size of the populations at time $t\ge 0$, their time evolution is supposed to be influenced by predation, growth and death processes. Therefore, this dynamics is classically described by the following pair of ordinary differential equations
\begin{equations} \label{LV}
\frac{d\, x(t)}{dt} &= \alpha x(t) -\beta \,x(t)y(t) ,\\
 \frac{d\,y(t)}{dt} &= - \delta\, y(t) + \gamma \,x(t)y(t) ,
\end{equations}
where the preys' population is characterized by the rate of Malthusian growth $\alpha\ge 0$ and by a predation rate $\beta\ge 0$. Furthermore, the predators' population is subject to losses due to death  with rate  $\delta\ge 0$ and on the growth due to predation defined by the parameter $\gamma\ge0$.

While less popular, the approach proposed in \cite{MBR,SR} is capable to establish an interesting parallel between prey-predator dynamics and economic modelling. This analogy, as documented by \cite{Gan}, goes back to the Italian economist Giuseppe Palomba, who first had used  Lotka-Volterra equations \fer{LV} in the 1939 book \cite{Pal}, and to Goodwin, who considered applications of Lotka-Volterra system to economy in the 1967 book \cite{Good}.

In this paper, we provide a kinetic interpretation of the introduced dynamics by means of a system of Boltzmann-type equations. The interacting species are characterized from a statistical perspective according to their size. Hence, the temporal evolution of the statistical distributions of two species, classically denoted as prey and predators,  is subject to variation due to birth and death as well as to interactions between the two families (predation). The building blocks of the model are given by the redistribution operator introduced  in \cite{BST}, which allows for a general representation of the birth and death process of the involved species, and a kinetic bilinear operator of Boltzmann type which quantifies the interaction between species. 
Interestingly, the structure of the interaction is similar to that recently introduced by the authors in connection with the study of kinetic models suitable for describing several patterns in compartmental epidemiology connected with the transition between susceptible and infected individuals \cite{DPaTZ, DPTZ}.

The linear operator introduced in \cite{BST} is sufficiently flexible to redistribute to agents a constant amount of wealth independently of the wealth itself, or to redistribute proportionally (or inversely proportionally) to their wealth. Hence, in the new application to a biological system, it allows the number of births  to be quantified in various ways and in terms of the size of the groups of the species.

In more details, the paper is organized as follows: in Section \ref{sect:2} we introduce a kinetic model describing the evolution of the sizes of groups of biological systems obeying a prey-predator dynamic. The transitions depend on elementary interactions determining the evolution in time of probability densities described by a system of non-Maxwellian Boltzmann-like equations. A redistribution operator depending only on the first two moments of the distribution is then considered. In Section \ref{sec:3} we recall basic results on the grazing limit in kinetic theory and we provide a Fokker-Planck description of the dynamics.  Finally, in Section \ref{sec:4} we provide several numerical tests to verify the consistency of the approach with the classical Lotka-Volterra model and we observe the emergence of fat-tailed distributions, in the group of preys, for suitable choices of the redistribution operator. 

\section{The model}\label{sect:2}

Following the general approach of \cite{PT} we consider a multiagent system composed by two species, identified as preys and predators,  that are characterized by their number of individuals $x\ge 0$ (respectively $y\ge0$). We denote by $f(x,t)$ (respectively $g(y,t)$) be the statistical distributions at a certain time $t \ge 0$ of the group of preys (respectively predators), classified in terms of their sizes $x,y \in \mathbb R_+$. 
Therefore, $f(x,t)dx$ indicates the fraction of the population preys with size in $[x,x + dx]$ at time $t\ge0$. Similarly, $g(y,t)dy$ is the fraction of the populations of predators characterized by a size in $[y,y+dy]$ at time $t\ge0$. Moreover, without loss of generality we fix the initial masses of the distributions $f$ and $g$ equal to one. Therefore, the mean sizes of the populations are given by 
\[
m_{f}(t) = \int_{\mathbb R_+}xf(x,t)dx,\qquad m_{g}(t) = \int_{\mathbb R_+}yg(y,t)dy.
\]
The goal is to characterize the time evolution of the pair of probability densities $f(x,t)$, $g(y,t)$ subject to changes including  births and deaths events, and obeying to objective mechanical and behavioral principles when interacting each other.

Let us first characterize the interaction rules between the two species. To achieve this, let us fix a certain time $t= t_0$, considered as the initial time of the observations,  and let $x_0,y_0 $ be the numbers of preys and predators in two groups at time $ t_0 $. Then, we evaluate the statistical distribution of the size of the groups at regular time intervals $ t_{n+1} $, with $n$ nonnegative integer, $ n\in \N_+$, consequent to  the elementary variations $x_n \to x_{n+1}$ and  $y_n \to y_{n+1}$ of the groups in each time interval. Since a prey-predator  interaction leads to a decrease in the number of preys, and, in the presence of a sufficiently high number of preys, to an increase of the number of predators, this change can be modeled as follows
\begin{equations}\label{coll1}
 x_{n+1} &= x_n  -  \Phi_\e(y_n) x_n +   \eta_\e(y_n) x_n. \\
 y_{n+1} &= y_n  + \Psi_\e(x_n) y_n +  \tilde\eta_\e(x_n)  y_n.
\end{equations}
Thus, at each time step, the number  of preys and predators is modified by two different mechanisms, expressed in mathematical terms by two multiplicative terms, both parameterized by a small positive parameter $\e \ll 1$, measuring the intensities of the single variations.

To clarify, the  function $\Phi_\e(\cdot)$ in the first equation of \fer{coll1} quantifies the predation rate, and is consistently assumed to be increasing with respect to the size of the predator population. Likewise, the function $\Psi_\e(\cdot)$ in the second equation of \fer{coll1} quantifies the growth of the predators, increasing with respect to the prey population at least when the number of preys is sufficiently consistent.  In reason of the previous remarks, in what follows we  fix 
\be\label{Phi}
\Phi_\e(y) =\e\beta \frac y{1+y}; \quad 
\Psi_\e(x) = \e \gamma \frac {x-\mu}{1+x}.
\ee
In \fer{Phi}, $\beta$, $\gamma$ and $\mu$ are positive constants. The constants $\beta$ and $\gamma$ measure the maximal predation rate, and, respectively, the maximal growth rate of predators. Last, the constant $\mu \ge 1$ measures the lowest size of the group of preys which is compatible with a growth of the group of predators. These constants are required to satisfy the bounds  
\be\label{bb}
\e\beta < 1; \quad \e\gamma\mu < 1.
\ee
These conditions ensure that the predation rate is always smaller than one, and that even in absence of preys, the number of predators does not collapse to zero in a single interaction.

In \fer{coll1}, random fluctuations of the groups due to unknown factors  in the environment, like immigration, natural events,  epidemics, etc.  are  expressed by  the independent random variables $\eta_\e(y), \tilde\eta_\e(x)$,  of zero mean and bounded variance, given by 
\be\label{var1}
\langle \eta_\e^2(y) \rangle  = \e\sigma\frac y{1+y}; \quad  \langle \tilde\eta_\e^2 (x)\rangle =    \e\tilde\sigma \frac {x}{1+x}. 
\ee
Note that, consistently with the deterministic variations, the variance of the random fluctuations of one class  is an increasing function of the size of the  other class.  

Starting from the elementary interactions \fer{coll1} we can easily write the kinetic equations describing the evolution in time of the pair of population densities. 
In a suitable scaling, the evolution of the densities  is
quantitatively described by a system of non-Maxwellian Boltzmann-like equations \cite{MTZ}
\begin{equation}
  \label{kine}
  \begin{split}
  \frac{\partial f}{\partial t} = Q_1(f,g), \\  \frac{\partial g}{\partial t} = Q_2(g,f),
  \end{split}
\end{equation}
where the bilinear operators on the right-hand side of both equations quantify the variations of the population densities due to the elementary interactions of type \fer{coll1} where the instantaneous microscopic  transitions $x\to x^\prime$ and $y\to y^\prime$ are defined as follows 
\[
\begin{split}
x^\prime &= x - \Phi_\epsilon(y)x + x \eta_\epsilon(y) \\
y^\prime &= y + \Psi_\epsilon(x)y+y\tilde\eta_\epsilon(x).
\end{split}
\]
The homogeneous Boltzmann equations \eqref{kine} can be fruitfully written in weak form. It corresponds to say that the
solution to \fer{kine} satisfies, for all smooth functions $\varphi$  the integro-differential equations
 \begin{equations}
  \label{kine-w}
\frac{d}{dt}\int_{\R_+}\varphi(x)f(x,t)\,dx &=
 \int_{\R_+^2} \kappa_1^\e(x,y)  \left\langle \varphi(x^\prime)-\varphi(x)\right\rangle
f(x,t) g(y,t) \,dx \, dy, \\ 
\frac{d}{dt}\int_{\R_+} \varphi(y)g(y,t)\,dy &=
 \int_{\R_+^2} \kappa_2^\e(x,y)  \left\langle \varphi(y^\prime)-\varphi(y)\right\rangle
f(x,t) g(y,t) \,dx \, dy.
 \end{equations} 
 In \fer{kine-w} the expression $\langle\cdot\rangle$ denotes mathematical expectation with respect to the random variables $\eta_\epsilon$, $\tilde \eta_\epsilon$, and the functions $\kappa_i^\e(x,y)$, $i =1,2$ are the interaction frequencies. In what follows, to maintain the connection between the kinetic model and the Lotka--Volterra system, we fix these frequencies in the form
 \be\label{fre}
 \kappa_1^\e(x,y) =\kappa^\e(y)= \frac{1+y}\e; \quad \kappa_2^\e(x,y) = \kappa^\e(x) = \frac{1+x}\e.
 \ee
 Within this choice, the frequency of interactions in the first equation grows with the number of predators, while in the second equation grows with the number of preys.
 
 By choosing $\varphi(x) =1$ in \fer{kine-w} we can easily observe that  $f(x,t)$ and $g(y,t)$ {remain probability densities} for any  time $t \ge0$, if they are so initially
 \be\label{m0}
\int_{\R_+} f(x, t)\,dx = \int_{\R_+} g(y,t)\,dy = 1 .
 \ee

Then, by choosing $ \varphi(x) = x$ in both equations \fer{kine-w} one easily obtains that the evolution of the mean values is defined by the coupled system
\be\label{mean2}
\begin{split}
\frac{dm_f(t)}{dt} &= -\beta \,m_f(t)m_g(t); \\
 \frac{dm_g(t)}{dt} &= - \gamma\mu\, m_g(t) + \gamma \,m_f(t)m_g(t).
\end{split}
\ee
Thus, if the functions $\Phi_\e$ and $\Psi_\e$ in \fer{coll1} are given by \fer{Phi}, and the collision frequencies in \fer{kine-w} are equal to \fer{fre}, the mean values of the densities satisfy the Lotka--Volterra system \fer{LV} in which the rate of birth $\alpha$ of the preys is set equal to zero.
{
 \begin{remark}  \emph{The closed evolution of the mean values, in the form of the Lotka--Volterra like system \fer{mean2} is closely dependent from the expression of the interaction coefficients and the random variations in \fer{coll1}, as well as from the interaction frequencies, as given by \fer{Phi}, \fer{var1} and \fer{fre}.  A different choice, while possible, will not lead in general to a closed expression of the evolution of the mean values, and the relationship with the Lotka--Volterra system is lost. }
\end{remark}
}

We may obtain a kinetic interpretation of the birth rate by complementing the kinetic system \fer{kine} with an additional operator  which accounts for the variation of densities due to births of individuals in the groups. As already discussed in the introduction, we will resort to the redistribution operator introduced in \cite{BST}. In this paper, wealth redistribution mimics the action of a government that use the proceeds of taxation to improve the financial situation of agents based on their income.

According to \cite{BST}, for a given probability density function $h(x,t)$, with mean $m(t)$, we  assume the birth operator of the form
 \be\label{redis}
 R_\chi^\alpha(h)(x,t) = \alpha \frac\partial{\partial x} \left[ \left(\chi x - (\chi + 1)m(t)\right)h(x,t)\right].
  \ee
 The weight factor multiplying the distribution function inside the square brackets in \fer{redis}  involves in the mechanism the moments of order zero and one. Such a weight function adds one real parameter $\chi$ to the dynamics. This new parameter characterizes the type of redistribution.  The parameter $\alpha>0$ is determined by the constraint that, while the redistribution operator preserves the total mass, the mean number of individuals increases by a factor $\alpha$. In fact, for each value of the constant $\chi\in \mathbb R$, we have 
 \be\label{res1}
  \int_{\R_+}x R_\chi^\alpha(h)(x,t) \, dx = \alpha\, m(t),
 \ee
and, provided $h(x,t)$ satisfies in addition the condition $h(0,t) = 0$, we also get
 \be\label{res0}
  \int_{\R_+}R_\chi^\alpha(h)(x,t) \, dx = 0.
 \ee
We can observe that the condition $h(0,t) = 0$ is not necessary in the special case $\chi = -1$, which reproduces an
anti-drift operator introduced by Slanina in  \cite{Sla} to mimic wealth redistribution effects. 

As shown in \cite{BST}, the effects of the operator $R_\chi^\alpha$ on the birth process on a group described by the density function $f$ are closely related to the values of the parameter $\chi$. The case in which the percentage of births is equally distributed independently of the size of the group  is achieved with the special option $\chi=0$. 
In all other cases the births are selective, and may correspond to some partition strategy.   {In the present case, as far as the preys are concerned,  it is suitable to take into account positive values of the parameter $\chi$, which correspond to impose that the percentage  of births is inversely proportional to the size of the group.  On the contrary, we assume $\chi=0$ for the predators group.}

Taking into account the previous discussion, the full system is obtained by adding to equations \fer{kine-w} the birth operators defined in \fer{redis}. Therefore, we have
 \begin{equations}
  \label{kine-wd}
\frac{d}{dt}\int_{\R_+}\varphi(x)f(x,t)\,dx = & \int_{\R_+}\varphi(x)R_\chi^\alpha(f)(x,t)\,dx \, +\\
+ &  \int_{\R_+^2} \kappa_1^\e(x,y) \left\langle \varphi(x^\prime)-\varphi(x)\right\rangle
f(x,t) g(y,t) \,dx \, dy, \\ 
\frac{d}{dt}\int_{\R_+}\varphi(y)g(y,t)\,dy =& \int_{\R_+}\varphi(y)R_{0}^\nu(g)(y,t)\,dy\, + \\
+ &  \int_{\R_+^2} \kappa_2^\e(x,y) \left\langle \varphi(y^\prime)-\varphi(y)\right\rangle
f(x,t) g(y,t) \,dx \, dy.
 \end{equations} 
In \fer{kine-wd}, the constant $\nu$ which defines the percentage of the mean number of births of the group of predators, is assumed to satisfy the bound $ \nu < \gamma\mu$. In this case, resorting to property  \fer{res1}, it is immediate to conclude that the evolution of the mean numbers $m_f(t)$ and $m_g(t)$ satisfies the Lotka-Volterra system \fer{LV}, where now $\delta = \gamma\mu -\nu >0$.

{
\subsection{Evolution of the moments}
System \fer{kine-wd} allows the calculation of the evolution in time of observable quantities. Among them,  the evolution of the principal moments, which is achieved in correspondence to the test functions $\varphi(r)= r^n, \, 0\le n\in \N$. In particular, by choosing $\varphi(x)= 1$, we see that the total mass is preserved in time.  The choice $\varphi(x)=x$ leads to the evolution of the mean values,  which satisfy the Lotka-Volterra system \fer{LV}, i.e.
\begin{equation}
\label{eq:mg_evo}
\begin{split}
\dfrac{d}{dt} m_f(t) &= \alpha m_f(t) - \beta m_f(t) m_g(t), \\
\dfrac{d}{dt} m_g(t) &= -\delta m_g(t) + \gamma m_f(t) m_g(t). 
\end{split}
\end{equation}
Then, the evolution of the variances
\be\label{var}
v_f(t) = \int_\R(x-m_f(t))^2f(x,t)\,dx, \quad v_g(t) = \int_\R(y-m_g(t))^2g(y,t)\,dy,
\ee
 is obtained by considering $\varphi(x) = (x-m_f(t))^2$ in the first equation of  \eqref{kine-wd} and $\varphi(y) = (y-m_g(t))^2$ in the second equation. For the preys equation we have
\begin{equation}
\label{eq:vf_coll}
\begin{split}
&\dfrac{d}{dt} \int_{\mathbb R_+}(x-m_f)^2 f(x,t)dx = \int_{\mathbb R_+}(x-m_f)R_\chi^\alpha(f)dx + \\
&\qquad\int_{\mathbb R_+^2} \dfrac{1+y}{\epsilon}\left\langle (x^\prime - m_f)^2 - (x-m_f)^2 \right\rangle f(x,t)g(y,t)dx\,dy =  \\
&(\sigma m_g - 2\alpha \chi -2\beta m_g)v_f + m_f^2\sigma m_g + \epsilon \beta^2\int_{\mathbb R_+^2}\dfrac{x^2y^2}{1+y}f(x,t)g(y,t)dx\,dy.
\end{split}
\end{equation}
Likewise, for the predators we obtain 
\begin{equation}
\label{eq:vg_coll}
\begin{split}
&\dfrac{d}{dt} \int_{\mathbb R_+}(y-m_g)^2 g(y,t)dy = \int_{\mathbb R_+}(y-m_g)R_0^\nu(g)dx + \\
&\qquad \int_{\mathbb R_+^2}\dfrac{1+x}{\epsilon} \left\langle(y^\prime-m_g)^2 - (y-m_g)^2 \right\rangle f(x,t)g(y,t)dx\,dy = \\
&(\tilde \sigma  + 2\gamma(m_f-\mu))v_g + \tilde \sigma m_fm_g^2 + \epsilon \gamma^2 \int_{\mathbb R_+} \dfrac{(x-\mu)^2y^2}{1+x}f(x,t) g(y,t)dx\,dy. 
\end{split}
\end{equation}
It is important to remark that equations  \eqref{eq:vf_coll}-\eqref{eq:vg_coll} contain terms which are not  principal moments. In the first equation, for example, this term is given by
\[
\int_{\mathbb R_+}\dfrac{y^2}{1+y}\, g(y,t)\,dy.
\]
While it can be easily bounded in terms of the moments of order zero and one,  it is not explicitly expressed in terms of them. 
Thus, at variance with the evolution of the mean values, as given by system \fer{eq:mg_evo}, the evolution of higher  moments can not be put in closed form. 
\
This characteristic is typical of kinetic equations which contain variable interaction frequencies, and it is reminiscent of the analogous situation encountered in classical kinetic theory of rarefied gases, where Maxwellian--type molecules are used  to avoid this problem \cite{Bob}.
}

\section{The limit of grazing predation rate}\label{sec:3}

We investigate now the situation in which most of the interactions between species correspond to a very small variation of the sizes of the groups, but at the same time it is possible to keep trace at the macroscopic level of all phenomena affecting predation rules. This kind of asymptotic analysis is referred to as \emph{quasi-invariant limit}, and it allows to prove that under proper assumptions on the interaction parameters,  the kinetic Boltzmann system may be approximated by a system of Fokker-Planck equations, allowing in various cases an analytical investigation of the shape of the solution. This limit corresponds to consider in \fer{kine-w} a Taylor expansion up to the second order of the differences 
\[
\varphi(x^\prime)-\varphi(x), \quad \varphi(y^\prime)-\varphi(y),
\]
and subsequently to consider $\e \to 0$, see \cite{PT,T} for further details. It can be proven that, in this asymptotic regime, the kinetic system \fer{kine-w} is well-approximated by the following system of  Fokker-Planck equations
 \begin{equations}
  \label{FP}
\frac{\partial f(x,t)}{\partial t} &= \frac{\partial}{\partial x}\Bigg\{ \frac{\sigma m_g(t)}2 \frac{\partial(x^2 f(x,t))}{\partial x}+ \\
 & +  \left[ (\beta m_g(t) +\alpha\chi) x - \alpha(\chi +1)m_f(t)\right]f(x,t)\Bigg\} \\
\frac{\partial g(y,t)}{\partial t} &= \frac{\partial}{\partial y}\Bigg\{\frac{\tilde\sigma m_f(t)}2 \frac{\partial(y^2 g(y,t))}{\partial y} + \\ & +  \left[ \gamma(\mu -m_f(t))y -\nu m_g(t)\right]g(y,t)\Bigg\}.
 \end{equations}
complemented by no-flux boundary conditions \cite{FPTT}. 
We can observe that the evolution of the mean values of \fer{FP} is consistent with the one obtained at the Boltzmann level  in \fer{kine-w}, so that the mean values $m_f(t)$, $m_g(t)$ satisfy the Lotka--Volterra system \fer{eq:mg_evo}. 

{Moreover, the structure of the Fokker--Planck system \fer{FP} allows the computation in closed form of the evolution of higher moments of its solutions.  For example, the evolution of the variances $v_f(t)$ and $v_g(t)$ defined in \fer{var}  is given by the system
\begin{equation}
\label{eq:varFP}
\begin{split}
\dfrac{d}{dt}v_f(t) &= [(\sigma - 2\beta) m_g(t) -2 \alpha \chi] v_f(t)+ \sigma m_g(t)m_f^2(t), \\
\dfrac{d}{dt}v_g(t) &= [(\tilde{\sigma}+2\gamma)m_f(t)-2\gamma\mu]v_g(t) + \tilde \sigma m_f(t)m_g^2(t),
\end{split}
\end{equation}
that coincide with equations \eqref{eq:vf_coll}-\eqref{eq:vg_coll} in the limit $\epsilon\to 0^+$. It is interesting to remark that, unlike the case of the mean values, the evolution of the variances depends on the diffusion coefficients $\sigma$ and $\tilde\sigma$ characterizing the randomness of the evolution, and from the parameter $\chi$ characterizing the birth operator \fer{redis}. 
In Figure \ref{fig:mvfg} we report the numerical evolution of coupled system for mean and variance defined by the equations \eqref{eq:mg_evo} and \eqref{eq:varFP}. We can observe how a kinetic version of the Lotka-Volterra model allows to obtain more information about the dynamics of the system of agents in the form of the evolution of higher order moments. 
\begin{figure}
\centering
\includegraphics[scale = 0.325]{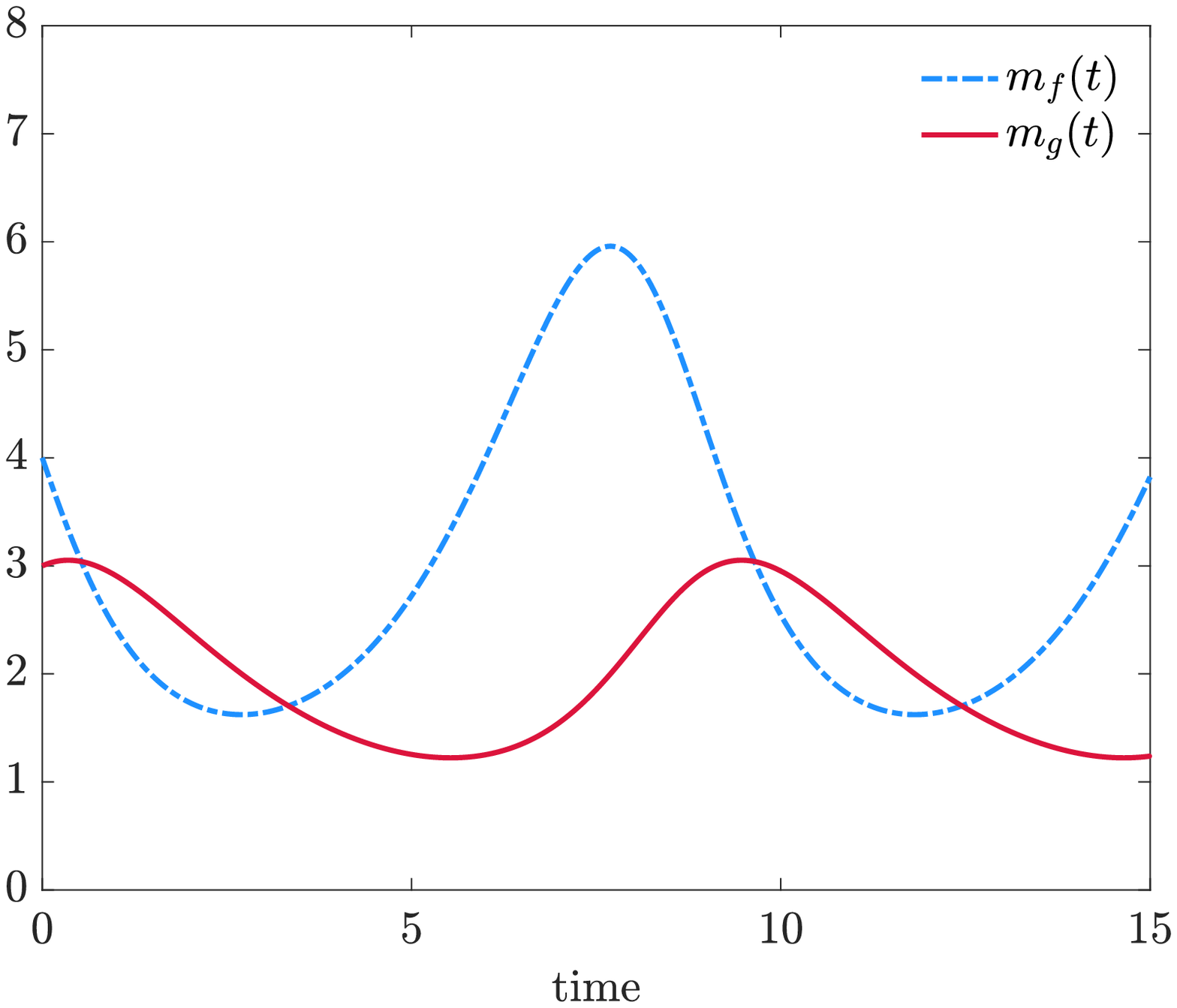}
\includegraphics[scale = 0.325]{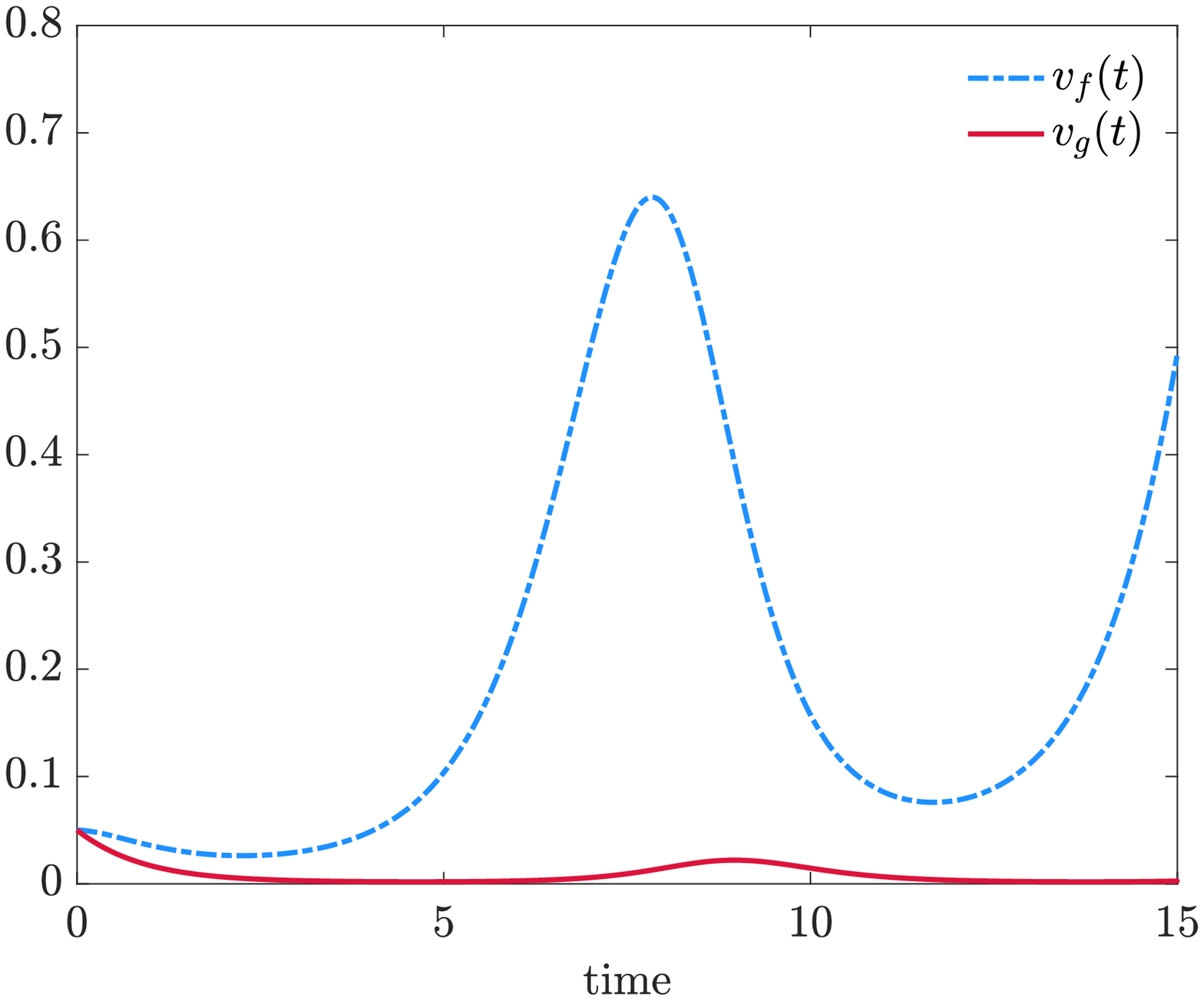}
\caption{Evolution of the coupled system \eqref{eq:mg_evo}-\eqref{eq:varFP} for the mean and variance obtained from the Fokker-Plank model \eqref{FP}. We considered $m_f(0) = 4$, $m_g(0) = 3$, $v_f(0) = v_g(0) = 0.05$. We fixed the parameter as in Table \ref{tab:1}.  }
\label{fig:mvfg}
\end{figure}
}


\section{Numerical tests}\label{sec:4}

In this section we provide several numerical tests to show the consistency of the proposed approach with the classical Lotka-Volterra dynamical system. Furthermore, we will show numerically how the kinetic approach allows additional information to be obtained on the large time behaviour of the densities of preys and predators in dependence of the parameters in the birth operators $R_\chi^\alpha(\cdot)$, $R_0^\nu(\cdot)$. For the approximation of the system of Fokker-Planck equations in \fer{FP} we will adopt the structure preserving numerical strategy developed in \cite{PZ}. These methods are capable to reproduce large time statistical properties of the solution density with arbitrary accuracy together with the preservation of its main physical properties, like its positivity.  We complement  the model \fer{FP}  with no-flux boundary conditions.

In all the following tests we will fix the parameter as reported in Table \ref{tab:1}.

\begin{table}
\begin{center}
\begin{tabular}{ |c| c| c| }
\hline
Parameter & Value & Meaning \\
\hline\hline  
 $\alpha$ & 1.00 & Growth rate preys \\ 
 $\beta$ & 0.50 & Predation rate preys \\  
 $\mu$ & 10.00 & Lowest size preys   \\
 $\gamma$ &0.15 & Growth rate predators \\
  $\nu$ & 1.00 & Birth rate predators\\
 $\delta$ & $\gamma\mu-\nu>0$ & Death rate predators\\
 $\sigma^2$ & $10^{-3}$ & Diffusion coefficient preys \\
 $\tilde \sigma^2$ & $10^{-3}$ & Diffusion coefficient predators \\
 \hline 
\end{tabular}
 \caption{Parameters chosen in the numerical tests. }
 \label{tab:1}
 \end{center}
\end{table}

\subsection{Test 1. Evolution of the mean values}

We consider the Fokker-Planck model \fer{FP} with initial distribution 
\begin{equation}\label{eq:fg0}
\begin{split}
f(x,0) = 
\begin{cases}
C_f \exp\left (-\dfrac{(x-m_f^0)^2}{2\kappa_f}\right) & x \ge 0 \\
0 & x<0
\end{cases} \\
g(y,0) = 
\begin{cases}
C_g \exp\left (-\dfrac{(y-m_g^0)^2}{2\kappa_g}\right) & y \ge 0 \\
0 & y<0,
\end{cases} 
\end{split}
\end{equation}
with $C_f,C_g>0$ normalization constants and with $m_f = 4$, $m_g = 3$ and $\kappa_f = \kappa_g = \frac{1}{10}$.

\begin{figure}
\centering
\includegraphics[scale = 0.325]{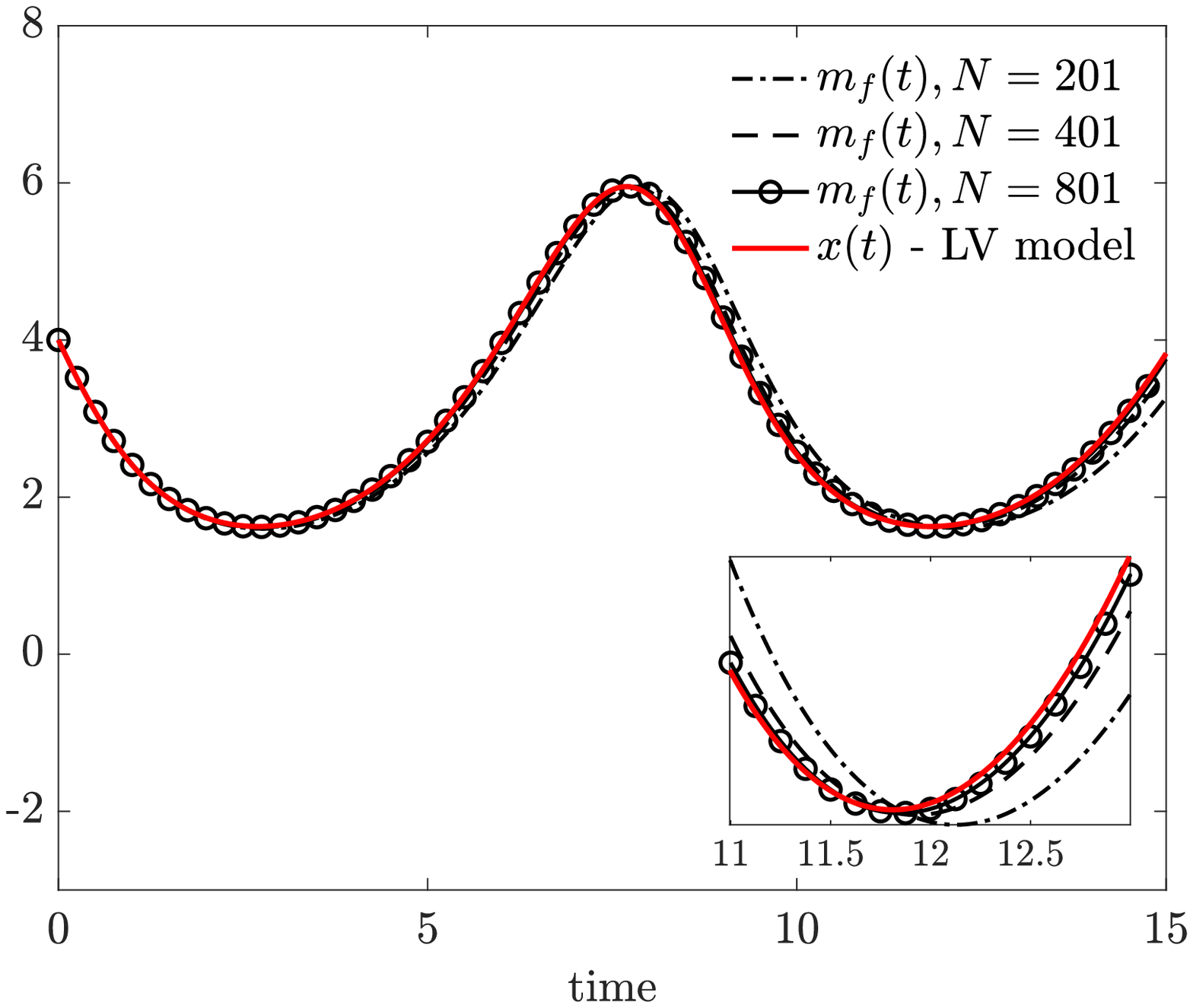}
\includegraphics[scale = 0.325]{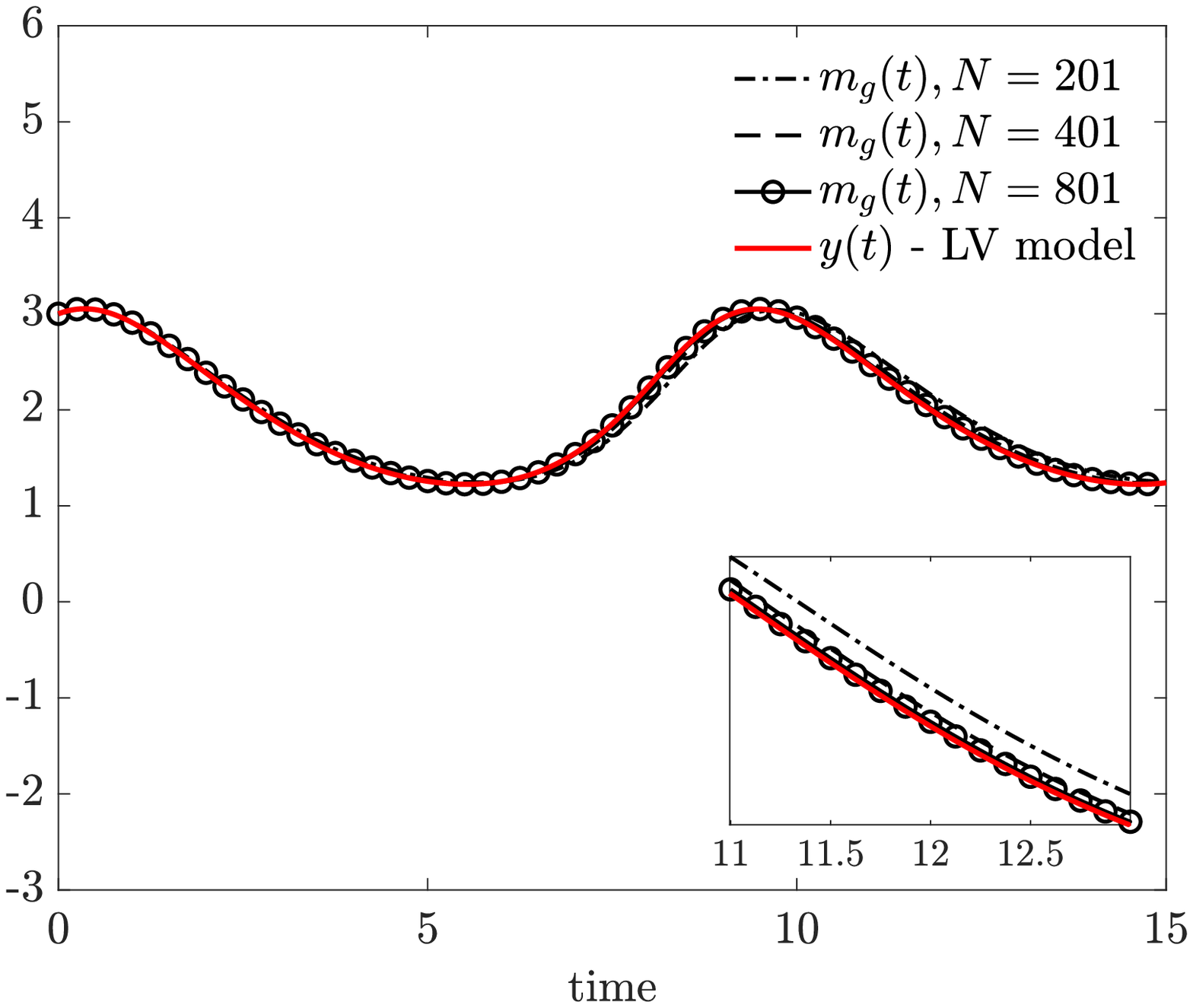}
\caption{\textbf{Test 1}. Comparison of the evolution of the mean values for the populations of preys $f(x,t)$ and predators $g(y,t)$ with the Lotka-Volterra dynamical system \fer{LV}. The Fokker-Planck system has been approximated by means of a 4th order SP methods with RK4 time integration and an increasing number of gridpoints $N = 201,401,801$ over the domain $[0,10]$. We considered $\Delta t = \Delta x^2= \Delta y^2$. }
\label{fig:evolution}
\end{figure}

In Figure \ref{fig:evolution} we show the evolution of the mean values $m_f$, $m_g$ of \fer{FP} over the time horizon $[0,15]$ for an increasing number of gridpoints $N = 201,401,801$ for the discretization of the interval $[0,L]$, $L= 10$. The choice of parameters has been defined in Table \ref{tab:1} and we further considered $\chi = 0$. We highlight in red the numerical solution of the Lotka-Volterra model obtained in the same regime of parameters. We can observe how, consistently with the analytical considerations in Section \ref{sec:3}, we can obtain numerically a good approximation of the original dynamical system in terms of the mean values of the distributions of preys $f(x,t)$ and predators $g(y,t)$. 

\subsection{Test 2. Emergence of fat-tailed distributions}

In this test we show the evolution of the approximated distributions $f(x,t)$ and $g(y,t)$ solution to \fer{FP} for varying values of the parameters $\chi \in \mathbb R$. All the other parameters have been chosen in agreement with Table \ref{tab:1}. We consider the initial distributions defined in \fer{eq:fg0} and we solve the system of Fokker-Planck equations \fer{FP} with $N = 801$ gridpoints in the interval $[0,10]$, and time frame $[0,15]$ with $\Delta t = \Delta x^2$. 

\begin{figure}
\centering
\includegraphics[scale = 0.325]{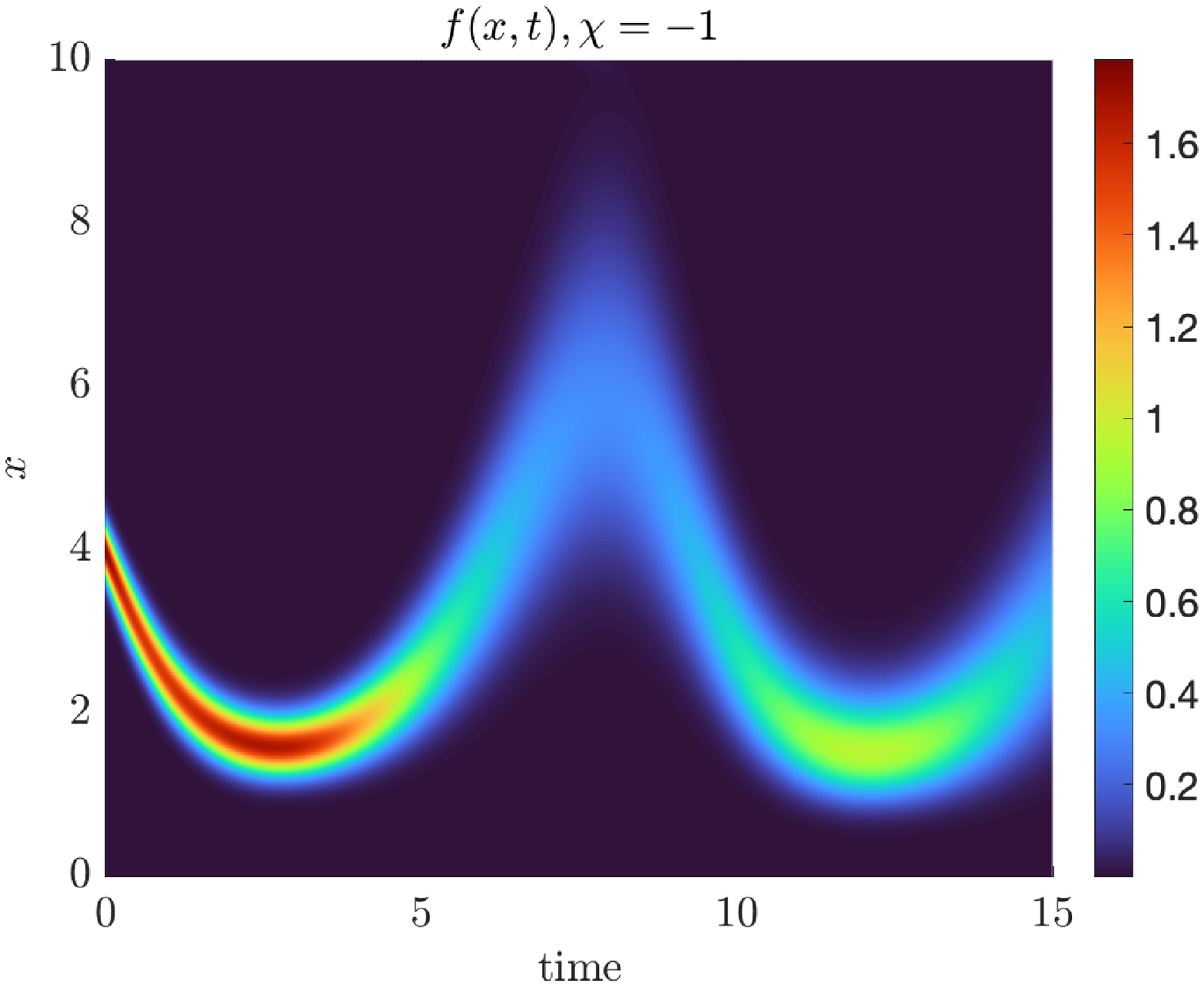}
\includegraphics[scale = 0.325]{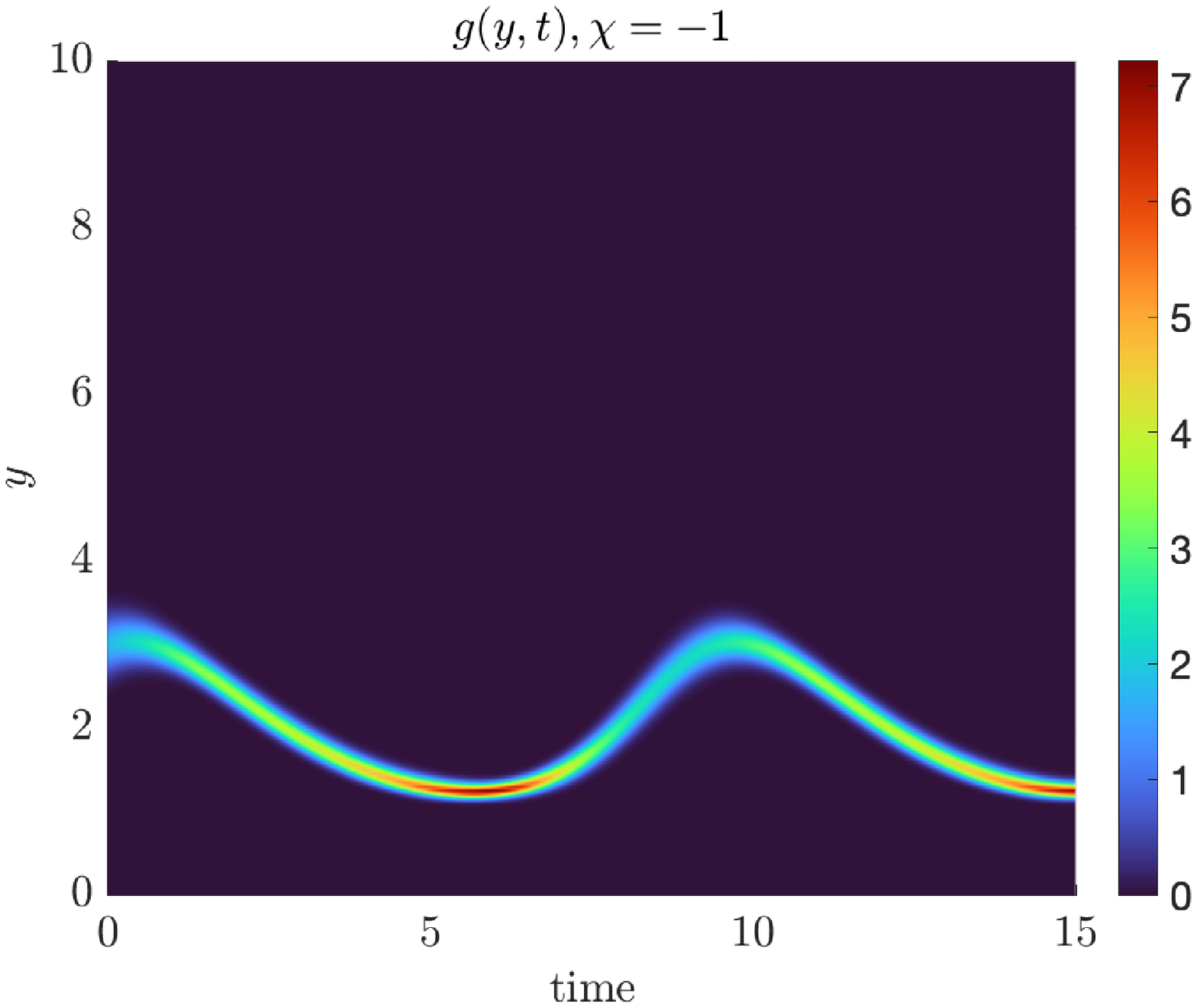} \\
\includegraphics[scale = 0.325]{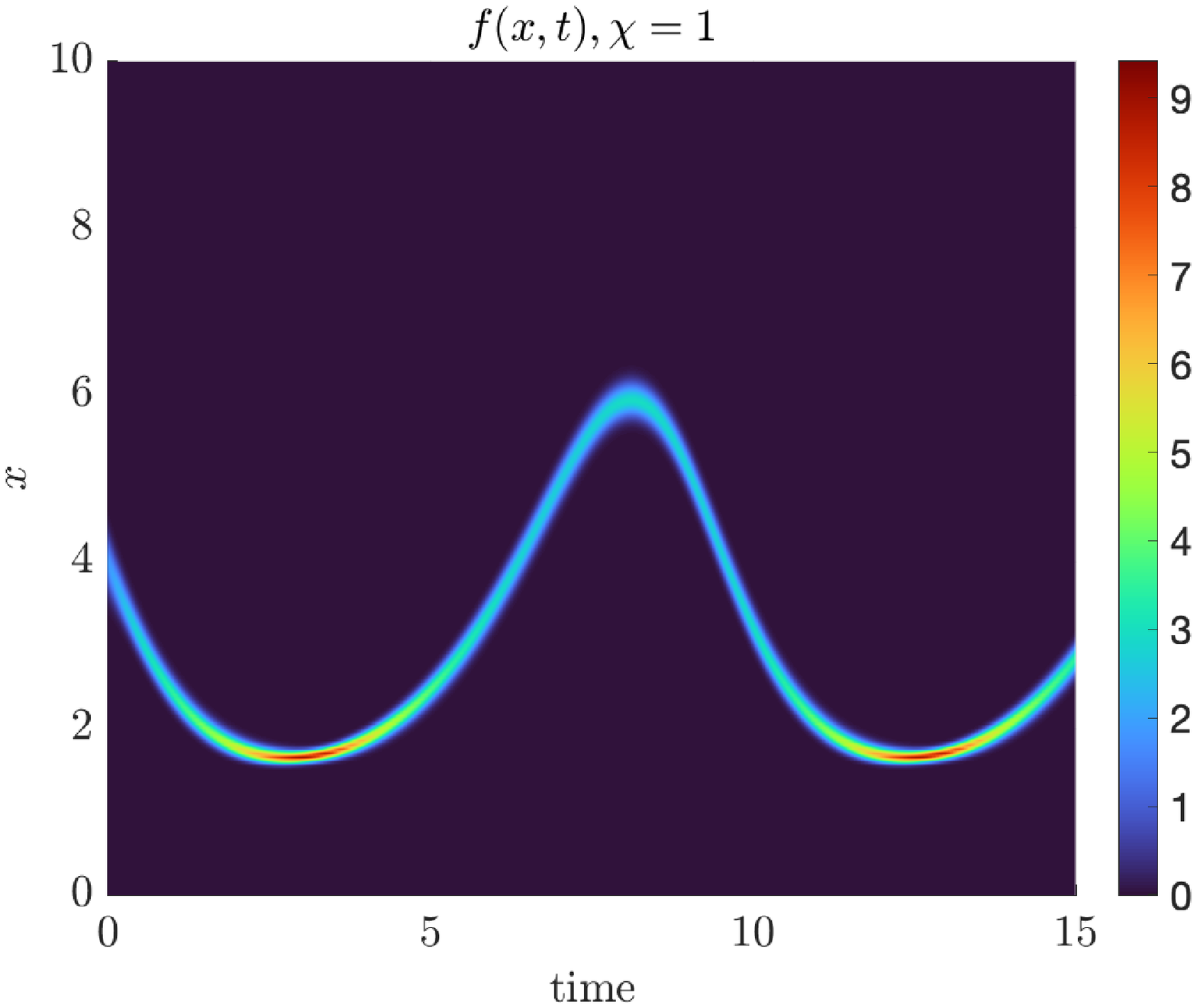}
\includegraphics[scale = 0.325]{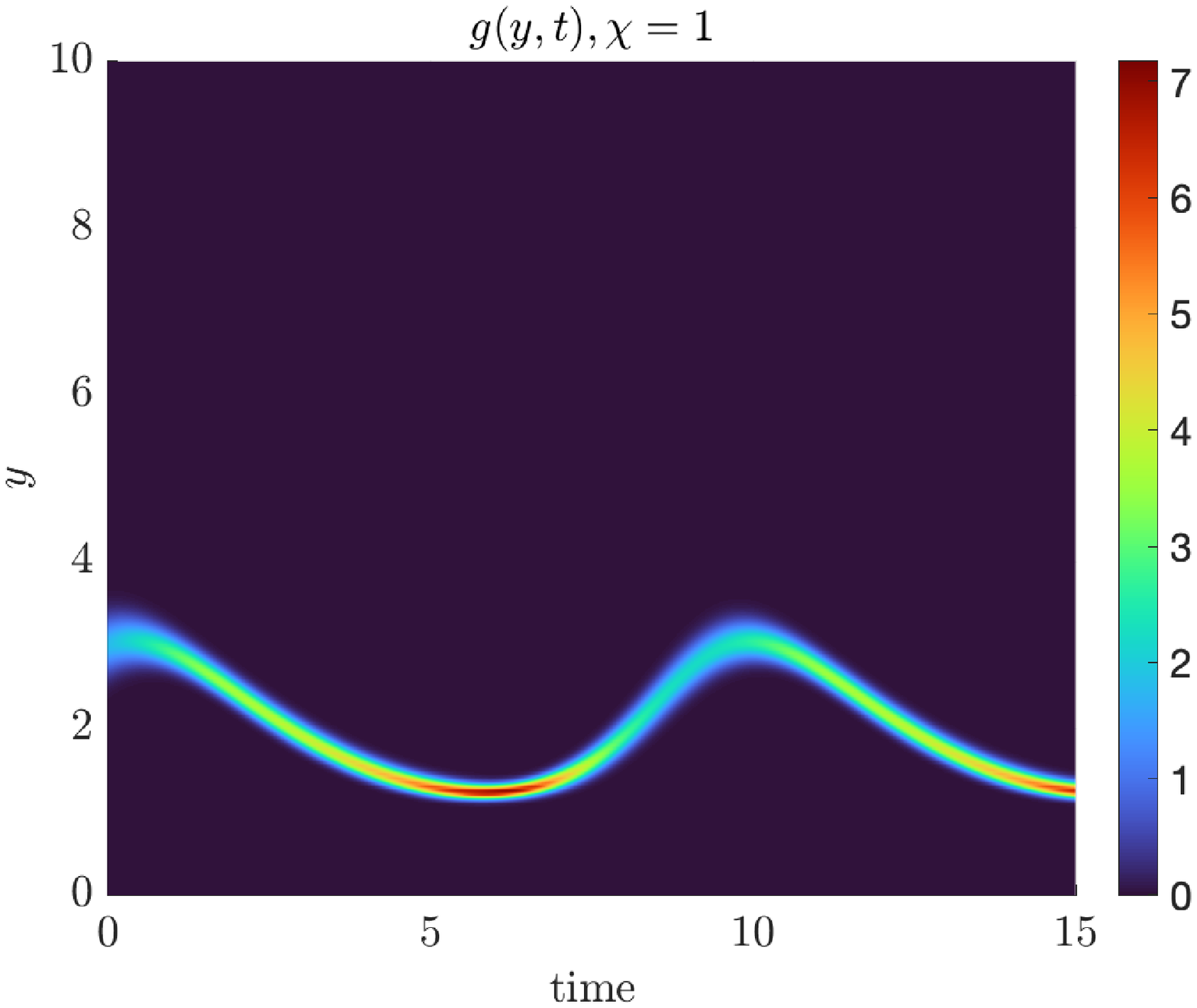} 
\caption{\textbf{Test 2}. Evolution of the distributions $f(x,t)$ and $g(y,t)$ over the time interval $[0,15]$ and for $x,y \in [0,10]$. Approximation of the system of Fokker-Planck equations obtained with 4th order SP scheme with RK4 time integration, $N = 801$ gridpoints and $\Delta t = \Delta x^2 = \Delta y^2$. We considered $\chi = -1$ (top row) or $\chi = 1$ (bottom row) in \fer{FP}.}
\label{fig:surf}
\end{figure}

In Figure \ref{fig:surf} we provide the total evolution of the distribution functions in the cases $\chi = -1$ (top row) and $\chi = 1$ (bottom row). As specified in Section \ref{sect:2} the two choices are coherent with two different redistribution policies. Indeed, for all $\chi >0$ the number of new births is distributed in small values of $x\ge 0$, whereas for $\chi<0$ the new births are concentrated in high values of $x \ge 0$. Coherently with \cite{BST}, we may observe how the preys population may exhibit a rather different behaviour for large times. Indeed, fat-tail-type distributions appears in the case $\chi = -1$. 

\begin{figure}
\includegraphics[scale = 0.325]{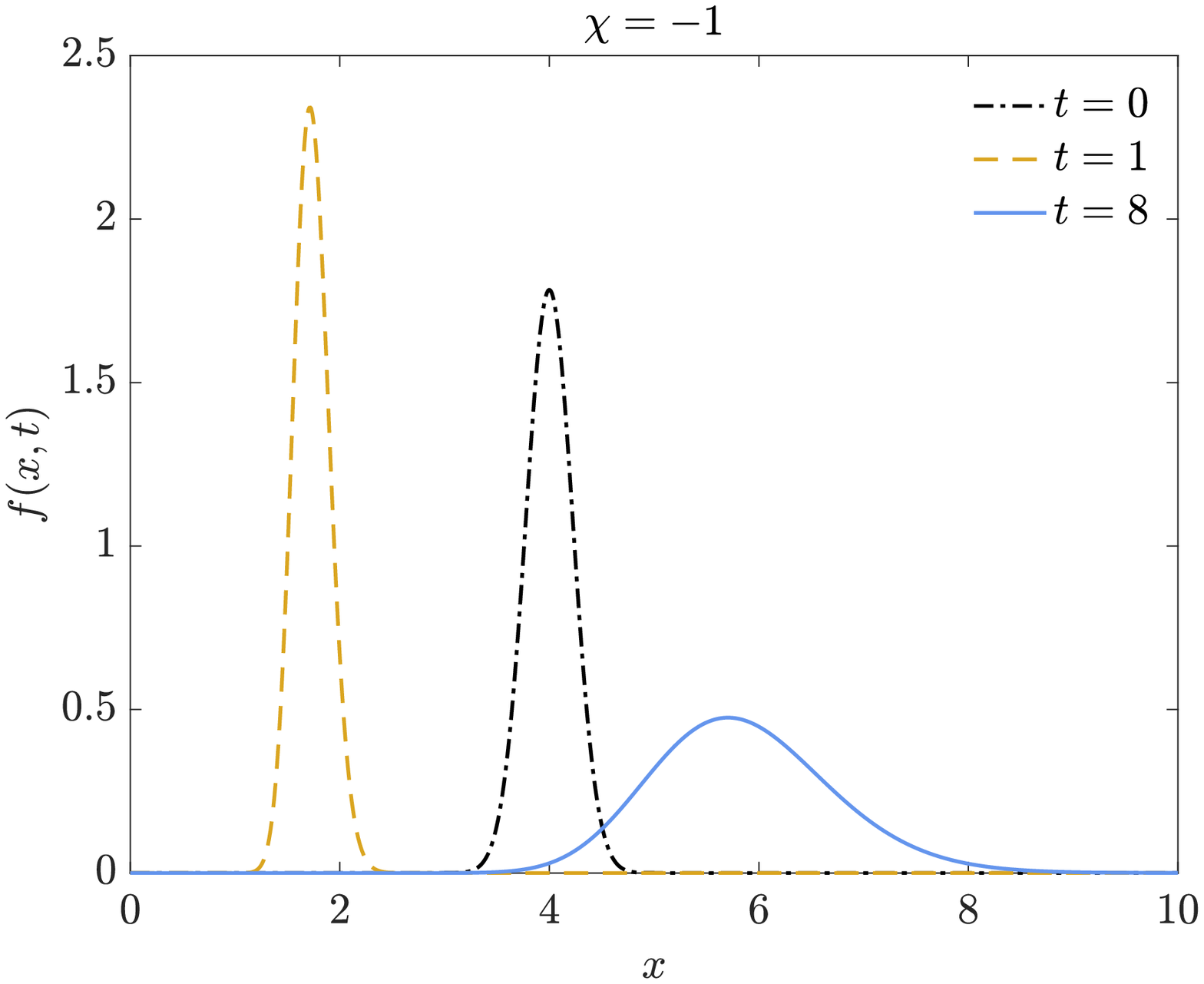}
\includegraphics[scale = 0.325]{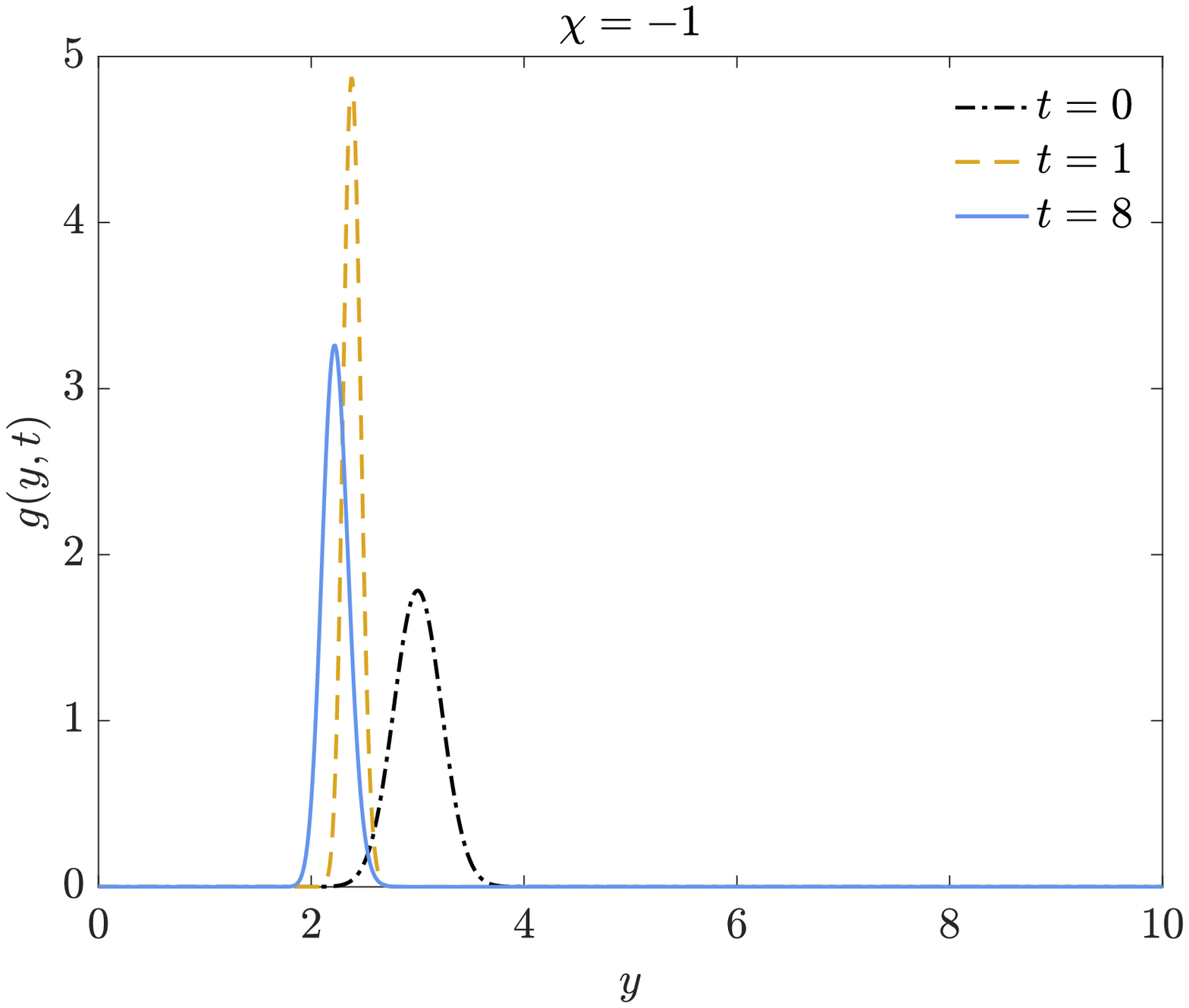}\\
\includegraphics[scale = 0.325]{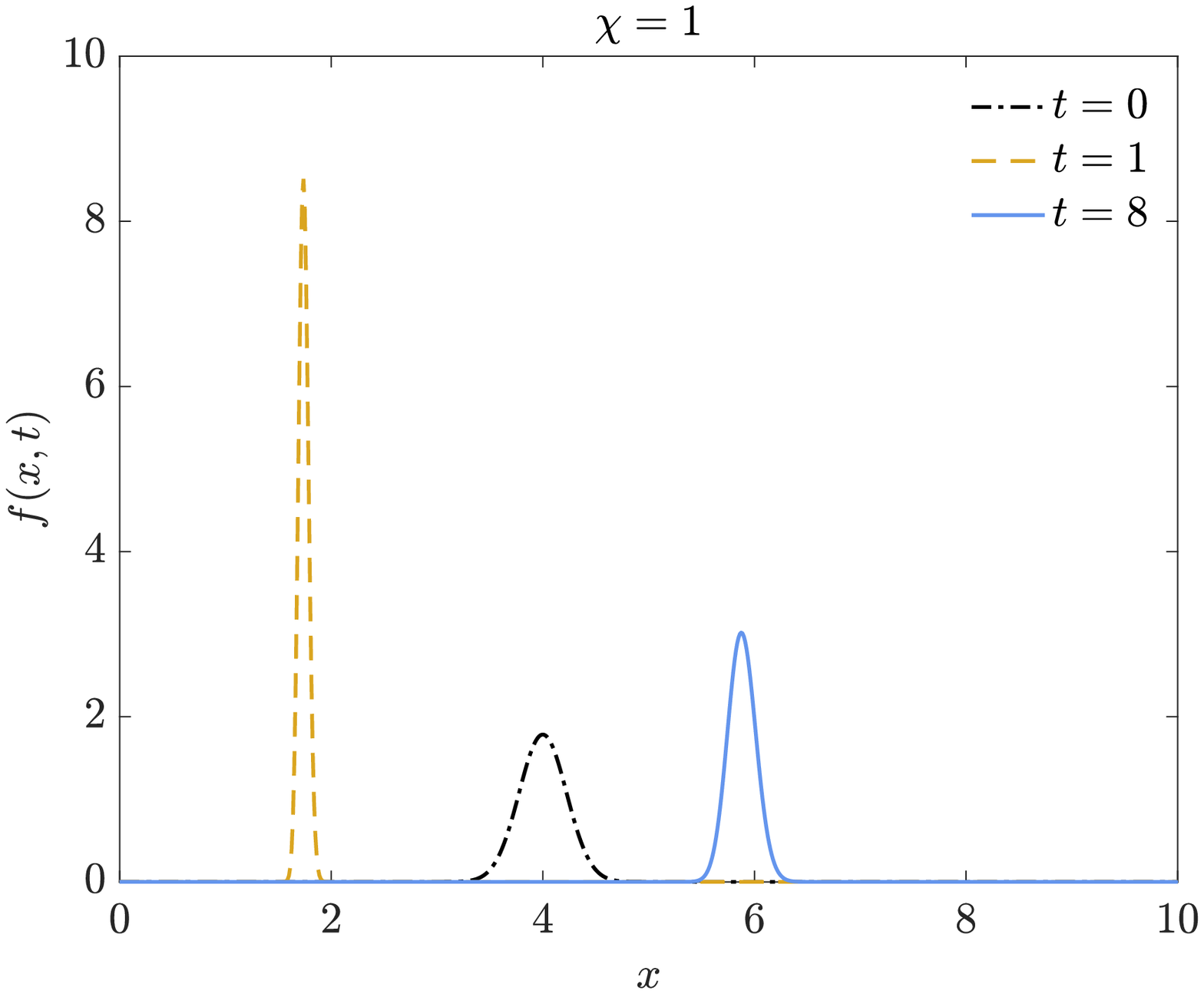}
\includegraphics[scale = 0.325]{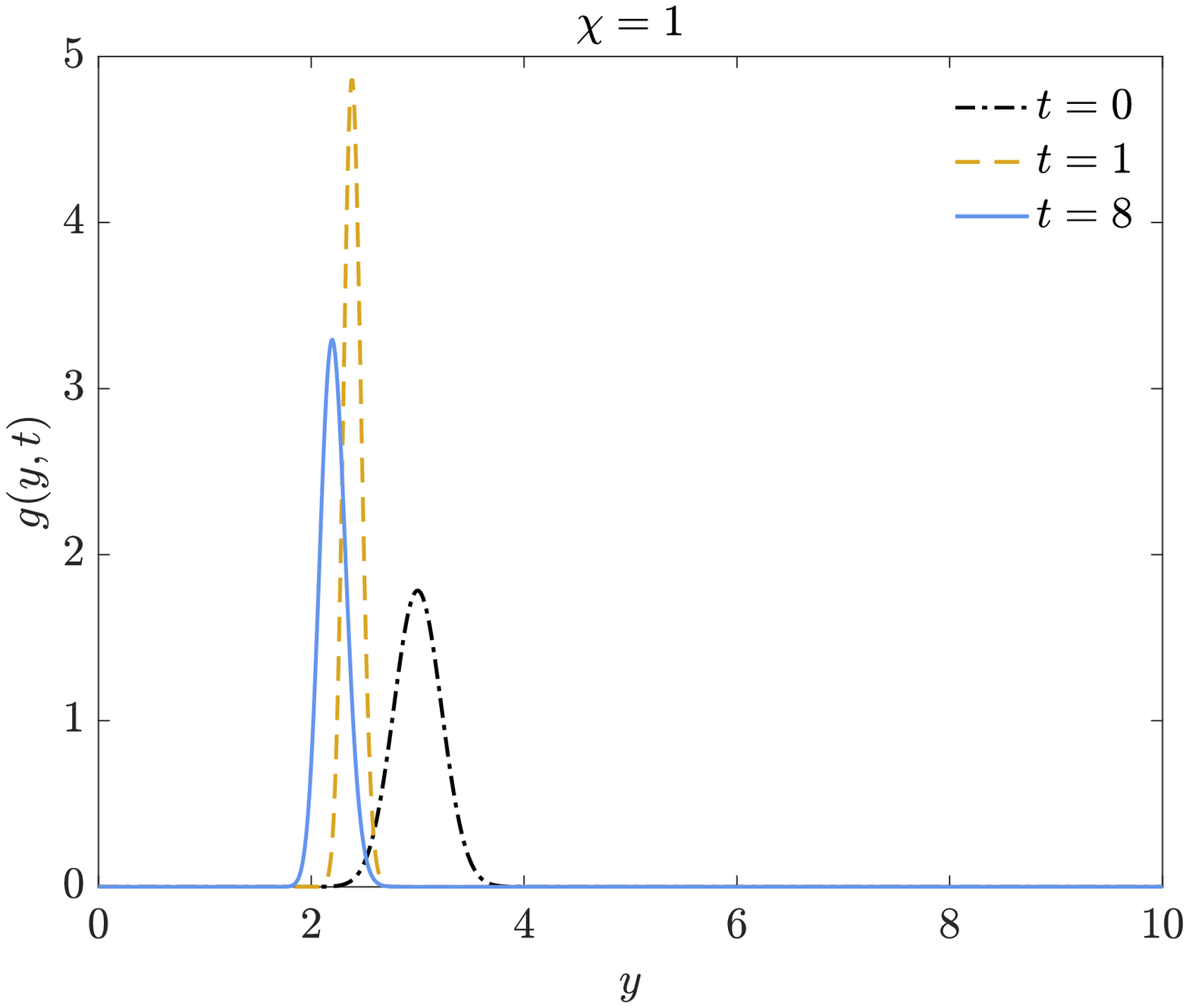}
\caption{\textbf{Test 2}. Distributions of preys $f(x,t)$ (left) and predators $g(y,t)$ (right) at three different time steps $t = 0,1,8$, $x,y \in [0,10]$. We considered $\chi = -1$ (top row) or $\chi = 1$ (bottom row) in \fer{FP}.}
\label{fig:3times}
\end{figure}

\begin{figure}
\centering
\includegraphics[scale = 0.325]{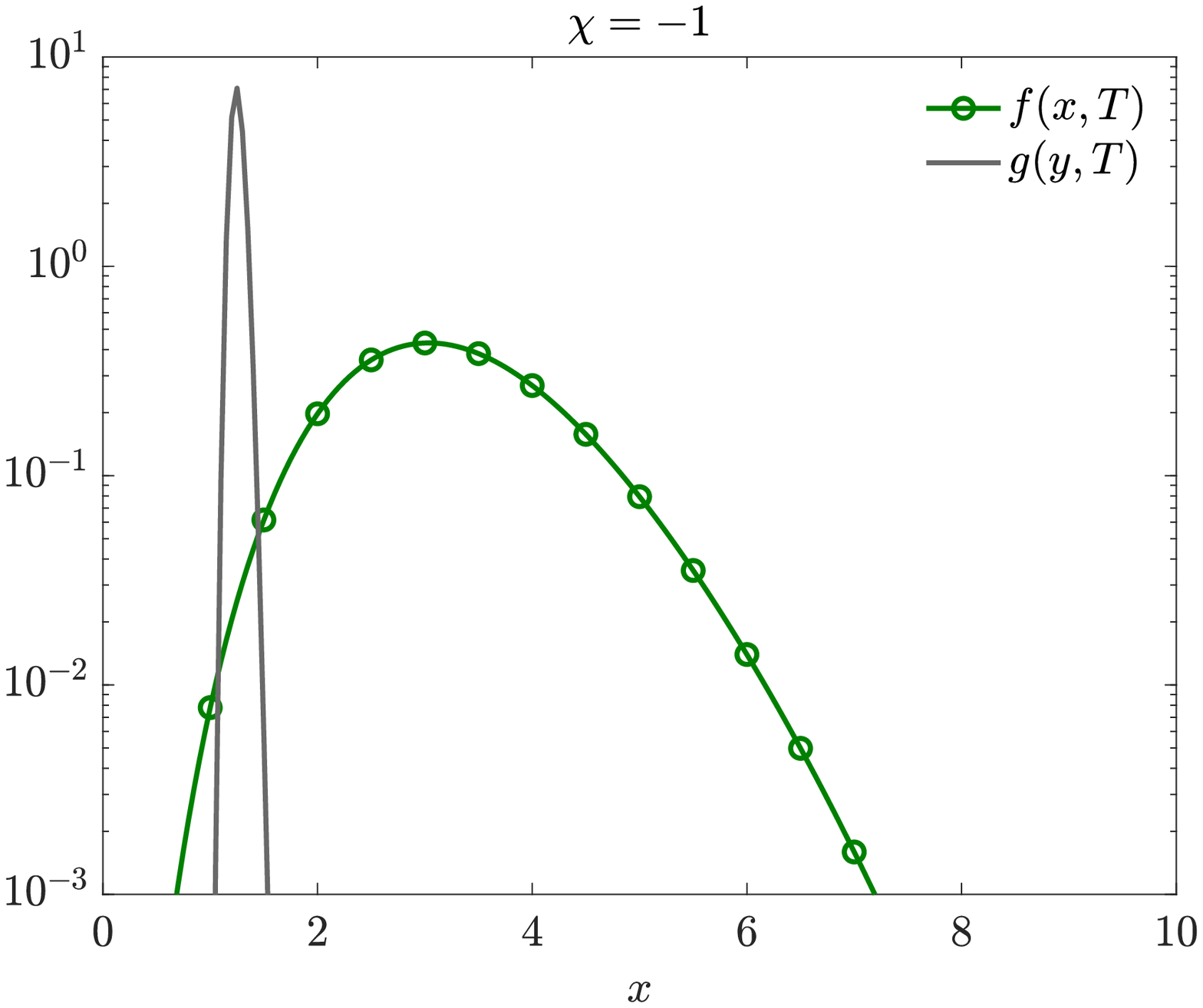}
\includegraphics[scale = 0.325]{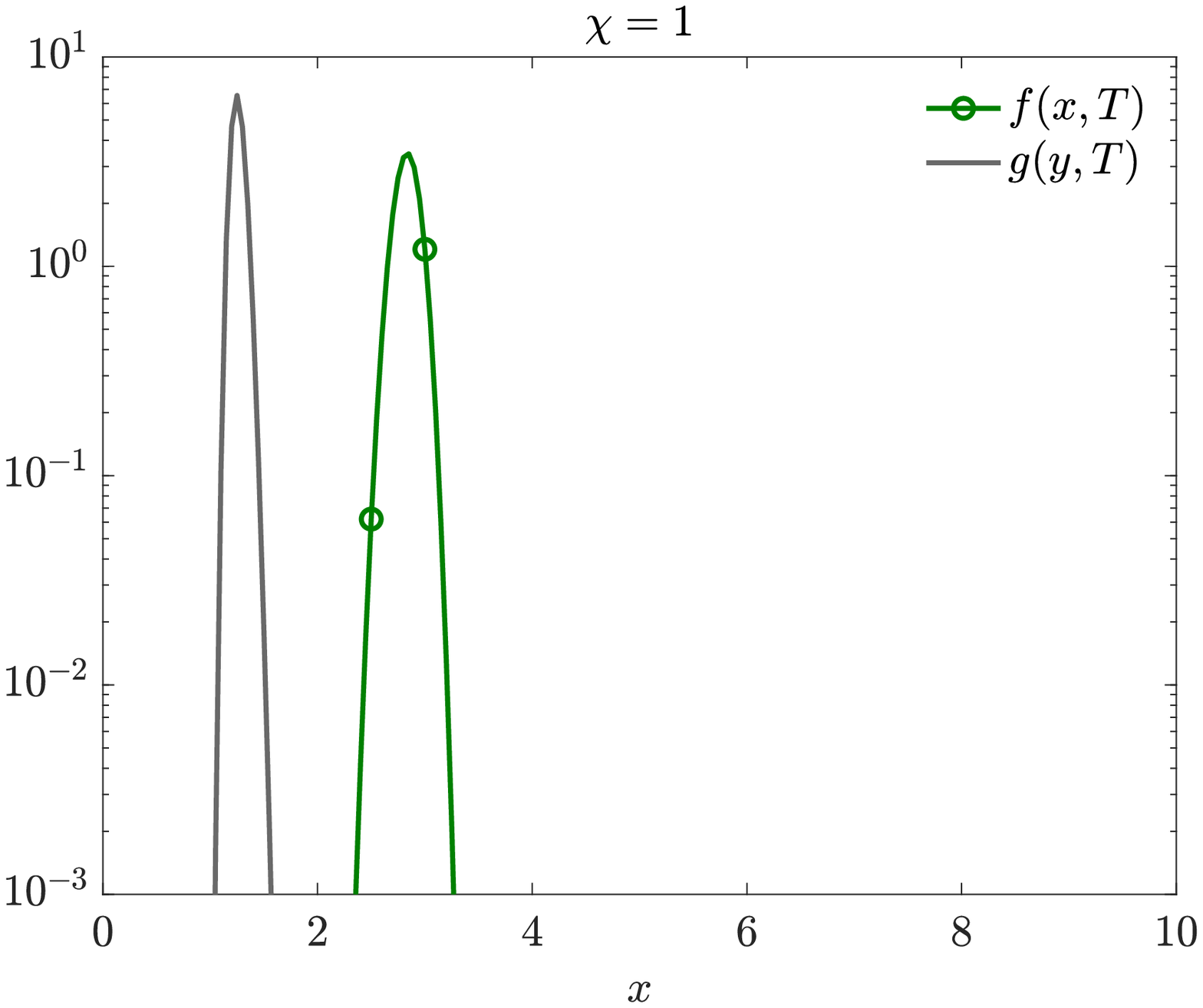}
\caption{\textbf{Test 2}. Distributions of preys $f(x,t)$ and predators $g(y,t)$ at time $T = 15$ in the cases $\chi = -1 $ (left) or $\chi = 1$ (right). }
\label{fig:loglog}
\end{figure}

To clarify this behaviour, we present in Figure \ref{fig:3times} a plot of the distributions of preys and predators at three fixed times $t = 0,1,8$. We can observe how for $\chi= -1$ the distribution of preys develops fat tails in finite time. This behaviour is further observable in loglog scale, see Figure \ref{fig:loglog}. 

\section{Conclusions}

Motivated by the similarities between the problem of wealth redistribution with taxation in a multi-agent society, we introduced in this paper a kinetic description of  the statistical distributions in time of the sizes of groups of biological systems obeying  a prey-predator dynamic. The kinetic system is built in such a way that the evolution of the mean values is driven by a classical Lotka-Volterra system of differential equations. Among other characteristics, it is shown that the time evolution of the probability distributions of the size of groups of the two interacting species is heavily dependent on the  kinetic redistribution operator quantifying the births. Numerical experiments clarify the time-behavior of the distributions of groups of the species, and highlight the possibility that the distribution of the sizes of preys could exhibit fat tails. {The system of Fokker--Planck equations \fer{FP} deserves a further analysis, which will be developed elsewhere. In particular, it would be of great interest to understand the role of the quasi-stationary solutions of the system, corresponding to the distributions annihilating the flux of \fer{FP}, and their relationship with  stability issues. }
 
 \vskip 1cm
\vspace{0.5cm} \indent {\it
A\,c\,k\,n\,o\,w\,l\,e\,d\,g\,m\,e\,n\,t\,s.\;} This work has been written within the activities of GNFM group of INdAM (National Institute of
High Mathematics).  MZ acknowledges the support of MUR-PRIN2020 Project No.2020JLWP23 (Integrated Mathematical Approaches to Socio-Epidemiological Dynamics). GT wish to acknowledge partial support by IMATI, Institute for Applied Mathematics and Information Technologies ``Enrico Magenes'', Via Ferrata 5 Pavia, Italy. He is grateful to the editors of this volume for granting the opportunity to collaborate in the remembrance of Giampiero Spiga, a friend, co-worker in many researches, and prominent figure in Italian Mathematical Physics.

\bigskip
\begin{center}

\end{center}

\bigskip
\bigskip
\begin{minipage}[t]{10cm}
\begin{flushleft}
\small{
\textsc{Giuseppe Toscani}
\\*University of Pavia,
\\*Department of Mathematics 
\\*Via Ferrata, 5
\\* 27100 Pavia, Italy
\\*e-mail: giuseppe.toscani@unipv.it
\\[0.4cm]
\textsc{Mattia Zanella}
\\*University of Pavia,
\\*Department of Mathematics 
\\*Via Ferrata, 5
\\* 27100 Pavia, Italy
\\*e-mail:mattia.zanella@unipv.it
\\[0.4cm]

}
\end{flushleft}
\end{minipage}


\begin{thebibliography}{99}
\frenchspacing \small





\bibitem{BST} {\sc   M. Bisi},  {\sc  G. Spiga} and   {\sc  G. Toscani}, \emph{Kinetic models of conservative economies with wealth redistribution}, Commun. Math. Sci. \textbf{7}(4), (2009) 901--916.

\bibitem{Bisi} {\sc M. Bisi} \emph{Some kinetic models for a market economy}, Boll. Unione Mat. Ital. \textbf{10}, (2017) 143--158.




\bibitem{Bob}  {\sc A.V. Bobylev}, \emph{The theory of the spatially Uniform Boltzmann equation for Maxwell molecules},
Sov. Sci. Review C \textbf{7}, (1988) 112--229.

 
\bibitem{Cha}  {\sc A. Chakraborti},  \emph{Distributions of money in models of market economy}, Int. J. Modern Phys. C
 \textbf{13}, (2002) 1315--1321.
 
 \bibitem{ChCh} {\sc A. Chakraborti} and   {\sc B.K. Chakrabarti}, \emph{Statistical mechanics of money: Effects of saving propensity}, Eur. Phys. J. B  \textbf{17}, (2000) 167--170.
 
\bibitem{CCM}  {\sc A. Chatterjee},  {\sc B.K. Chakrabarti} and  {\sc S.S. Manna}, \emph{Pareto law in a kinetic model of market with random saving propensity}, Physica A   \textbf{335}, (2004) 155--163.

\bibitem{CYC}   {\sc A.Chatterjee},  {\sc S.Yarlagadda} and   {\sc B.K.Chakrabarti} Eds. \emph{Econophysics of wealth distributions},
New Economic Window Series, Springer-Verlag (Italy), 2005.

\bibitem{CCS}   {\sc A. Chatterjee},  {\sc B.K. Chakrabarti} and   {\sc R.B. Stinchcombe}, \emph{Master equation for a kinetic model of
trading market and its analytic solution},  Phys. Rev. E  \textbf{72}, (2005) 026126.


\bibitem{CPT}  {\sc S. Cordier},  {\sc L. Pareschi} and   {\sc G. Toscani}, \emph{On a kinetic model for a simple market economy}. J. Stat. Phys.  \textbf{120},   (2005) 253--277.

\bibitem{DLR}
{\sc P. Degond}, {\sc J.-G. Liu} and {\sc C. Ringhofer}. \emph{Evolution of the distribution of wealth in an economic environment driven by local Nash equilibria}. J. Stat. Phys., \textbf{154} (2014) 751--780. 

\bibitem{DPaTZ} {\sc G. Dimarco}, {\sc  L.Pareschi},   {\sc G. Toscani} and  {\sc M. Zanella}, \emph{Wealth distribution under the spread of infectious diseases}  Phys. Rev. E  \textbf{102}, (2020) 022303.

\bibitem{DPTZ} {\sc G. Dimarco}, {\sc  B.Perthame},   {\sc G. Toscani} and  {\sc M. Zanella}, \emph{Kinetic models for epidemic dynamics with social heterogeneity}, J. Math. Biol. \textbf{83}, (2021) n.4.

 \bibitem{DY}   {\sc A. Dragulescu} and  {\sc V.M. Yakovenko}, \emph{Statistical mechanics of money}, Eur. Phys. Jour. B  \textbf{17},  (2000). 723--729.
 
 \bibitem{DMT}   {\sc B. D\"uring},  {\sc D. Matthes} and  {\sc G. Toscani}, {Kinetic equations modelling wealth redistribution: A comparison  of approaches}, Phys. Rev. E,  \textbf{78}, (2008)  056103.

\bibitem{DMT2}  {\sc B. D\"uring},  {\sc D. Matthes} and  {\sc G. Toscani},  \emph{A Boltzmann type approach to the formation of wealth distribution curves}, Riv. Mat. Univ. Parma, Riv. Mat. Univ. Parma \textbf{8}(1),  (2009) 199--261.

\bibitem{FPTT} {\sc G. Furioli}, {\sc A. Pulvirenti}, {\sc E. Terraneo} and {\sc G. Toscani}, \emph{Fokker--Planck equations in the modelling of socio-economic phenomena}, Math. Models Meth. Appl. Sci. \textbf{27}(1), (2017) 115--158.

\bibitem{FPTT2}
{\sc G. Furioli}, {\sc A. Pulvirenti}, {\sc E. Terraneo} and {\sc G. Toscani}, \emph{Non-Maxwellian kinetic equations modeling the evolution of wealth distribution}, Math. Mod. Meth. Appl. Sci. \textbf{30}(4), (2020) 685--725. 

\bibitem{Gan} {\sc G. Gandolfo}, \emph{Giuseppe Palomba and the Lotka-Volterra equations}, Rend. Fis. Acc. Lincei  \textbf{19},  (2008) 347--357.
 
 \bibitem{Good} {\sc R.M.Goodwin},   \emph{A Growth Cycle}, In C.H. Feinstein (ed.). \emph{Socialism, Capitalism and Economic Growth}. Cambridge University Press, Cambridge 1967.

\bibitem{Hay}   {\sc B. Hayes},  \emph{Follow the money}, American Scientist  \textbf{90},  (2002) 400--405.

 \bibitem{IKR}  {\sc S. Ispolatov},   {\sc P.L. Krapivsky} and   {\sc S. Redner}  \emph{Wealth distributions in asset exchange models}, Eur.
Phys. Jour. B \textbf{2},  (1998) 267--276.

\bibitem{MBR}  {\sc O. Malcai},  {\sc O. Biham},  {\sc P. Richmond} and   {\sc S. Solomon}, \emph{Theoretical analysis and simulations of the
generalized Lotka--Volterra model}, Phys. Rev. E  \textbf{66}, (2002)  031102.

 \bibitem{MT}  {\sc D. Matthes} and  {\sc G. Toscani},  \emph{On steady distributions of kinetic models of conservative economies},
J. Stat. Phys.  \textbf{130}, (2008) 1087--1117.

\bibitem{MTZ}
{\sc A. Medaglia}, {\sc A. Tosin}, and {\sc M. Zanella}, \emph{Monte Carlo stochastic Galerkin methods for non-Maxwellian kinetic models of multiagent systems with uncertainties}, Part. Diff. Equat. Appl. \textbf{3}, (2022) Art. n. 51.

\bibitem{Murray}
{\sc J. D. Murray}, \emph{Mathematical Biology I: An Introduction.}, Interdisciplinary Applied Mathematics, Springer New York 2002. 

\bibitem{Pal} {\sc G. Palomba}, \emph{Introduzione allo studio della dinamica economica}, Napoli: Jovene, Napoli (1939).

\bibitem{PT}  {\sc L. Pareschi} and  {\sc G. Toscani}, \emph{ Interacting Multiagent Systems: Kinetic Equations and Monte Carlo Methods}, Oxford University Press, Oxford (2014).

\bibitem{PZ}
{\sc L. Pareschi} and {\sc M. Zanella}, \emph{ Structure preserving schemes for nonlinear Fokker-Planck equations and applications}, J. Sci. Comput., \textbf{74}, (2018) 1575--1600. 

\bibitem{T}
{\sc G. Toscani}, \emph{Kinetic models of opinion formation}, Commun. Math. Sci. \textbf{4}(3), (2006) 481--496.

\bibitem{TTZ}
{\sc G. Toscani}, {\sc A. Tosin} and {\sc M. Zanella}, \emph{Kinetic modelling of multiple interactions in socio-economic systems}, Netw. \& Heter. Media \textbf{15}(3), (2020) 519--542. 

\bibitem{Sla}  {\sc F. Slanina},  \emph{Inelastically scattering particles and wealth distribution in an open economy}, Phys. Rev. E  \textbf{69}, (2004)  046102.

 \bibitem{SR} {\sc S. Solomon} and {\sc P. Richmond},  \emph{Stable power laws in variable economies; Lotka--Volterra implies Pareto--Zipf} Eur. Phys. J. B  \textbf{27}, (2002)  257--262.

\end{thebibliography}
\end{document}